\documentclass[twocolumn,preprintnumbers,amsmath,amssymb]{revtex4}
\usepackage{epsfig}
\usepackage{color}
\usepackage[colorlinks,urlcolor=blue,linkcolor=blue,citecolor=red]{hyperref}
\begin{document}
\setlength{\voffset}{1.0cm}
\title{Untwisting twisted NJL$_2$-kinks by a bare fermion mass}
\author{Michael Thies\footnote{michael.thies@gravity.fau.de}}
\affiliation{Institut f\"ur  Theoretische Physik, Universit\"at Erlangen-N\"urnberg, D-91058, Erlangen, Germany}
\date{\today}

\begin{abstract}
Twisted kinks in the massless NJL$_2$ model interpolate between two distinct vacua on the chiral circle. If one approaches the 
chiral limit from finite bare fermion masses $m_0$, the vacuum is unique and twist cannot exist. This issue 
is studied analytically in the non-relativistic limit, using a no-sea effective theory.
We conclude that even in the massless limit, the interpretation of the twisted kink has to be revised. One has to attribute the fermion number of
the valence state to the twisted kink. Fermion density is spread out over the whole space due to the massless pion field.
The result can be pictured as a composite of a twisted kink (carrying energy, but no fermion number) and a partial winding of the chiral spiral  
(carrying fermion number, but no energy). This solves at the same time the puzzle of missing baryons with fermion number $N_f<N$
in the massless NJL$_2$ model.
\end{abstract}
\maketitle
\section{Introduction}
\label{sect1}
The 1+1 dimensional, massless Gross-Neveu (GN) models \cite{1} with discrete or continuous chiral symmetry are well studied
integrable quantum field theories.
These are interacting theories of $N$ flavors of massless Dirac fermions with point-like four-fermion interactions,
\begin{eqnarray}
{\cal L}_{\rm GN} & = &  \bar{\psi} i \partial \!\!\!/  \psi + \frac{g^2}{2} (\bar{\psi}\psi)^2 ,
\nonumber \\
{\cal L}_{{\rm NJL}_2} & = & \bar{\psi}  i \partial \!\!\!/ \psi + \frac{g^2}{2}\left[  (\bar{\psi}\psi)^2 + (\bar{\psi} i \gamma_5 \psi)^2 \right] .
\label{A1}
\end{eqnarray} 
Flavor indices are suppressed as usual. The version with U(1) chiral symmetry will be referred to as Nambu--Jona-Lasinio model \cite{2} in
two dimensions (NJL$_2$). We are interested in the large $N$ limit of 't~Hooft \cite{3}, keeping $Ng^2$ constant, where semi-classical
methods can be applied reliably. Many results have been obtained in closed, analytical form in the past, of interest to both particle and condensed matter 
physics. 

If we add a bare mass term 
\begin{equation}
\delta {\cal L} = -  m_0 \bar{\psi}\psi
\label{A2}
\end{equation}
to the Lagrangians (\ref{A1}), chiral symmetry is explicitly broken and 
integrability is lost. The response of the system to such a term depends very much on the symmetry. In the case of the GN model with discrete chiral symmetry, 
it has been known for some time that exact static solutions (baryons, multi-baryon bound states) can be taken over from the
massless model after a mere redefinition of certain parameters \cite{4,5,6}. This does not work anymore for time dependent problems, like breathers or baryon-baryon
scattering \cite{7}. Recently, time dependence 
was studied near the non-relativistic limit \cite{8}, using a microscopic ``no-sea" effective theory where negative energy
states are projected out \cite{9}. It was found that to leading order in the non-relativistic expansion, the massive GN model remains
integrable with only forward elastic scattering. To next order (``fine structure" type corrections) however, backward
elastic scattering and inelastic processes set in. To establish this fact requires the numerical solution of partial differential
equations (PDE's), in contrast to the analytical results in the massless case.

In the present work, we turn our attention to the massive NJL$_2$ model. Here, even static baryons are hard to come by
analytically. The only case where a systematic analytical approximation scheme has been found
is the region of complete filling (fermion number $N$) close to the chiral limit. As first shown by Salcedo et al. \cite{10},
here the baryon problem can be mapped onto the sine-Gordon equation, incidentally exactly as in QCD$_2$. 
Systematic corrections for larger bare masses have been computed using derivative expansion methods \cite{5}.
Specific for these solitonic baryons is the fact that baryon number arises as winding number of a pseudoscalar
boson field analogous to the pion in QCD, 
strongly reminiscent of the Skyrme model \cite{11} in 3+1 dimensions. This enables one
to completely avoid the solution of Hartree-Fock (HF) equations and work exclusively with bosonic fields. Apart from this,
there exist some numerical investigations \cite{12}, but analytical insight into the near-chiral limit regime for fermion numbers between 0 and $N$ is
still painfully missing.

Here, we propose to address this problem near the non-relativistic fermion limit, using the no-sea effective theory.
In contrast to Ref.~\cite{12} where something complementary was done, we are interested in the region very close to the 
chiral limit. In order to make progress, we shall restrict ourselves to fermion numbers $N_0/N \ll 1$ where
a non-relativistic approach can be trusted. No matter how small the bare mass is, the
vacuum is unique and ``twisted kinks" joining two arbitrary points on the chiral circle do not exist.
It is in this sense that bare mass must ``untwist" twisted kinks.
We would like to see how -- if at all -- twisted kinks are approached when coming from 
the massive fermion side. In this way, we hope to shed some new light
on the nature of the somewhat exotic twisted kinks. 

This paper is organized as follows. In Sect.~\ref{sect2}, we briefly review twisted NJL$_2$ kinks. 
The no-sea effective theory, our main tool, will be introduced in Sect.~\ref{sect3}. We then proceed to
reduce the Dirac equation to the non-relativistic Schr\"odinger equation in Sect.~\ref{sect4} and to
formulate and solve the twisted kink problem in this framework, Sect.~\ref{sect5}. After these 
preparations, we can proceed to the massive NJL$_2$ model and its non-relativistic limit in
Sect.~\ref{sect6}. Sect.~\ref{sect7} shows how one can solve the untwisted kink problem analytically
to a good approximation, near the chiral limit. In Sect.~\ref{sect8} we put together the insights gained 
here with what is already known about NJL$_2$ baryons with maximal fermion number and 
sketch a new picture of twisted kinks and their fermion content. We finish with a brief summary and
conclusions, Sect.~\ref{sect9}.

\section{Reminder of twisted NJL$_2$-kinks}
\label{sect2}

\begin{figure}
\begin{center}
\epsfig{file=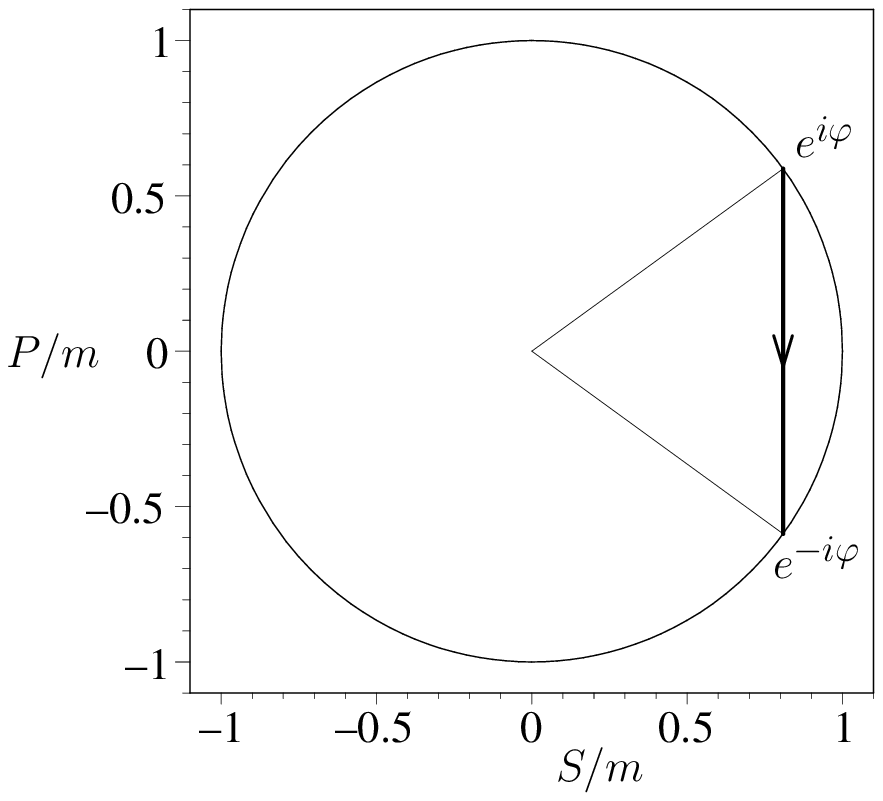,width=8cm,angle=0}
\caption{Trajectory of a twisted kink in the ($S,P$)-plane. The kink lives on the chord joining the two vacua 
with chiral angles $\pm \varphi$. The arrow points into the direction of increasing $x$.}
\label{fig1}
\end{center}
\end{figure}

In the large $N$ limit, the method of choice for studying bound state and scattering problems is the
relativistic version of the time dependent Hartree-Fock (TDHF) approximation. The 
mean-field Dirac equation of the NJL$_2$ model reads
\begin{equation}
\left( i  \partial \!\!\!/ - S -i \gamma_5 P \right) \psi = 0 
\label{B1}
\end{equation}
where the scalar ($S$) and pseudoscalar ($P$) potentials are related to the fermion spinors via the 
self-consistency conditions
\begin{eqnarray}
S & = &  - g^2 \langle \bar{\psi} \psi \rangle = - g^2 \sum_{\alpha}^{\rm occ} \bar{\psi}_{\alpha} \psi_{\alpha},
\nonumber \\
P & = &  - g^2 \langle \bar{\psi} i \gamma_5 \psi \rangle = - g^2 \sum_{\alpha}^{\rm occ} \bar{\psi}_{\alpha} i \gamma_5 \psi_{\alpha}.
\label{B2}
\end{eqnarray}
The sum over occupied orbits includes the Dirac sea and the occupied positive energy bound states, hereafter referred to as valence levels. 
The homogeneous solution describing the vacuum is not unique, but characterized by a chiral vacuum angle $\theta$,
\begin{equation}
\Delta = S-iP = m e^{i \theta}.
\label{B3}
\end{equation}
The U(1) manifold of all possible vacua is called the chiral circle. Its radius is the dynamical fermion mass $m$, generated
by dimensional transmutation from a dimensionless coupling constant as encoded in the vacuum gap equation
\begin{equation}
\frac{\pi}{Ng^2} = \ln \frac{\Lambda}{m}.
\label{B4}
\end{equation}
The TDHF solution of interest here is the twisted kink first found by Shei with inverse scattering theory \cite{13}.
In the ($S,P$) plane, it traces out 
a chord between two arbitrary points on the chiral circle.  By a proper choice of the global chiral angle, it can be cast into the form
\begin{equation}
\Delta = m \frac{e^{i\varphi} + e^{-i \varphi} e^{2\xi}}{1 + e^{2\xi}} = m \left( \cos \varphi - i \sin \varphi \tanh \xi \right)
\label{B5}
\end{equation}
with $\xi = m x \sin \varphi$ in the rest frame. The potential $\Delta$ interpolates between two vacua at $\theta= \varphi$ ($x \to - \infty$) and $\theta = - \varphi$
($x \to \infty$), see Fig.~\ref{fig1}. The kink potential has a single bound state which can be filled with $N_0\in [0,N]$ fermions. Self-consistency
relates the filling fraction $\nu=N_0/N$ (a continuous parameter in the large $N$ limit) to the ``twist angle" $\varphi$ 
\begin{equation}
\nu = \frac{\varphi}{\pi}.
\label{B6}
\end{equation}
The corresponding valence fermion density, 
\begin{equation}
\rho_{\rm val} = N_0 \frac{m \sin \varphi}{2 \cosh^2 \xi},
\label{B7}
\end{equation}
is cancelled exactly by the density induced in the Dirac sea (i.e., arising from the negative energy continuum states), so that the fermion density vanishes
identically. This is a consequence of axial and vector current conservation in the massless NJL$_2$ model, specific to 1+1 dimensions \cite{14}.
The energy eigenvalue of the valence state is $E=m \cos \varphi$, crossing the whole mass gap as a function of filling fraction.
The mass of the twisted kink is $\frac{Nm}{\pi} \sin \varphi$. Twist angles $\varphi$ and $\pi -\varphi$ yield the same mass and correspond  
to kink and anti-kink, joining the two vacua $e^{\pm i \varphi}$ in opposite direction. Two or more twisted kinks can form bound states or undergo
scattering, and there are twisted breathers in addition. All of these processes can be explicitly calculated by analytical means
thanks to the integrability of the massless NJL$_2$ model \cite{15,16}.
Here, we shall be mainly concerned with the single twisted kink at rest and its fate under switching on a bare mass term.

\section{No-sea effective theory}
\label{sect3}

In view of a later application to the massive NJL$_2$ model which is not exactly solvable, we focus on the non-relativistic
regime where some analytical insight can be obtained. For the kink at rest considered here, this means a restriction to 
small twist angle $\varphi$ (or, equivalently, small occupation fraction $\nu$). In this case, the spatial structure of the kink is smooth
and the binding energy small on the scale of the fermion mass $m$.
We can then take advantage of a ``no-sea effective theory" developed before \cite{9}. In the present section, we sketch its salient features for the massless 
NJL$_2$ model, postponing the massive model to later sections. 

In this effective theory, one eliminates the Dirac sea and works with
positive energy fermion states (valence levels) only. The effects of the Dirac sea are taken into account through modifications 
of the Lagrangian. Technically, this has only been possible in the non-relativistic regime so far. The effective Lagrangian derived in Ref.~\cite{9} 
to leading order (LO) in the non-relativistic expansion reads
\begin{eqnarray}
{\cal L}_{\rm eff} & = & \bar{\psi} ( i \partial \!\!\!/ - m ) \psi + \frac{\pi}{2N} (\bar{\psi}\psi)^2 + \frac{1}{2} \partial_{\mu} \Pi \partial^{\mu} \Pi
\nonumber \\
& & + \sqrt{\frac{4\pi}{N}} m \bar{\psi}i \gamma_5 \psi \Pi + \frac{2\pi m}{N} \bar{\psi}\psi \Pi^2.
\label{C1}
\end{eqnarray}
The derivation of Eq.~(\ref{C1}) is quite subtle due to infrared problems generated by the massless pion field which necessitate re-summations.
The result is a field theory of interacting non-relativistic fermions and a pseudoscalar boson, the ``pion" field $\Pi$. The classical Euler-Lagrange equations
are
\begin{equation}
\left( i \partial \!\!\!/ - m + \frac{\pi}{N} \bar{\psi}\psi + \sqrt{\frac{4\pi}{N}} m i \gamma_5 \Pi + \frac{2\pi m}{N} \Pi^2 \right) \psi  =  0,
\label{C2}
\end{equation}
\begin{equation}
\left( \square - \frac{4\pi m}{N} \bar{\psi}\psi \right) \Pi  =  \sqrt{\frac{4\pi}{N}} m \bar{\psi}i \gamma_5 \psi.
\label{C3}
\end{equation}
In the large $N$ limit, we can again use mean field methods. The pion field becomes classical, whereas the (bound state) fermions satisfy the TDHF
equation (\ref{B1}), but now with the following self-consistency conditions
\begin{eqnarray}
S & = & m - \frac{\pi}{N} \langle \bar{\psi}\psi \rangle_+ - \frac{2\pi m}{N} \Pi^2,
\nonumber \\
P & = & - \sqrt{\frac{4\pi}{N}} m \Pi,
\label{C4}
\end{eqnarray}
\begin{equation}
\left( \square - \frac{4\pi m}{N} \langle \bar{\psi}\psi \rangle_+ \right) \Pi   =   \sqrt{\frac{4\pi}{N}} m \langle \bar{\psi}i \gamma_5 \psi \rangle_+.
\label{C5}
\end{equation}
Here, condensates with subscript $+$ involve only the (occupied) valence levels
\begin{eqnarray}
\langle \bar{\psi}\psi \rangle_+ & = &  \sum_{\alpha}^{\rm val} \bar{\psi}_{\alpha} \psi_{\alpha},
\nonumber \\
\langle \bar{\psi}i \gamma_5\psi \rangle_+ & = &  \sum_{\alpha}^{\rm val} \bar{\psi}_{\alpha} i \gamma_5\psi_{\alpha}.
\label{C6}
\end{eqnarray}
An interesting observable in the effective theory is the fermion density. In Ref.~\cite{9} it was shown that after elimination of the Dirac sea, the ``induced fermion density" 
gets transferred into the pion field, much like in the Skyrme model, 
\begin{equation}
\rho_{\rm ind} = \sqrt{\frac{N}{\pi}} \partial_x \Pi = - \frac{N}{\pi} \frac{\partial_x P}{2m}.
\label{C7}
\end{equation}
Hence the total fermion density assumes the following form 
\begin{equation}
\rho = \rho_{\rm val} + \rho_{\rm ind} = \sum_{\alpha}^{\rm val} \psi_{\alpha}^{\dagger} \psi_{\alpha}  + \sqrt{\frac{N}{\pi}}\partial_x \Pi.
\label{C8}
\end{equation}
In Ref.~\cite{9}, this theory has been formulated for Dirac fermions. Its validity is restricted to non-relativistic fermions however, whereas the pion remains relativistic. 
It is therefore desirable to combine the no-sea effective theory with a non-relativistic reduction of the Dirac equation to a Schr\"odinger equation, to the 
same LO that underlies the effective Lagrangian. This will be done in the following section. 

\section{Non-relativistic reduction of the Dirac equation}
\label{sect4}
Here, we carry out the non-relativistic reduction of the Dirac-TDHF equation in the massless NJL$_2$ model. Unlike in the GN model, only the LO Lagrangian is available in the no-sea effective theory.
Therefore it is sufficient to also work out the non-relativistic reduction to LO only (no ``fine structure" corrections \cite{8}). 
Starting point is the Dirac-TDHF equation (\ref{B1}). Using the Dirac-Pauli representation of the $\gamma$ matrices 
($ \gamma^0 = \sigma_3, \gamma^1 = i \sigma_2, \gamma_5 = \sigma_1$), pulling out the fast factor $e^{-imt}$ from the spinors and using the Hamiltonian form
of the Dirac equation, we arrive at
\begin{equation}
i \partial_t \left( \begin{array}{c} \psi_1 \\ \psi_2 \end{array} \right) = \left( \begin{array}{cc} S-m & -i \partial_x +i P \\ -i \partial_x - i P & -S-m \end{array} \right)
\left( \begin{array}{c} \psi_1 \\ \psi_2 \end{array} \right).
\label{D1}
\end{equation}
Next we eliminate the ``small component" $\psi_2$ (to LO in the non-relativistic expansion only),
\begin{equation}
\psi_2  =  - \frac{i}{2m} (\partial_x+P) \psi_1 .
\label{D2}
\end{equation}
This yields the following Schr\"odinger equation for the ``large component" $\psi_1$
\begin{equation}
i \partial_t \psi_1  =    
\left( - \frac{\partial_x^2}{2m} + S - m - \frac{(\partial_x P)}{2m} + \frac{P^2}{2m} \right) \psi_1.
\label{D3}
\end{equation}
The potentials ($S,P$) have been given above, see Eq.~(\ref{C4}). If we plug them in, the $P^2$-term in Eq.~(\ref{D3}) cancels the $\Pi^2$-term in $S$, Eq.~(\ref{C4}),
with the result
\begin{equation}
i \partial_t \psi_1 = \left( - \frac{\partial_x^2}{2m} - \frac{\pi}{N}  \langle \bar{\psi}\psi \rangle_+ - \frac{(\partial_x P)}{2m} \right) \psi_1.
\label{D4}
\end{equation}
Guided by the exact one-soliton solution, we assume the following dependence on the expansion parameter $\epsilon$ to LO
\begin{eqnarray}
\psi_1 & \sim & \epsilon^{1/2}, \quad \psi_2 \sim \epsilon^{3/2}, \quad S-m \sim \epsilon^2, 
\nonumber \\
P & \sim & \epsilon, \quad \partial_x  \sim  \epsilon, \quad \partial_t \sim \epsilon^2.
\label{D5}
\end{eqnarray}
To this order, the condensates become
\begin{eqnarray}
\langle \bar{\psi} \psi \rangle_+ & = &  \sum_{\ell} N_{\ell} |\psi_{1,\ell}|^2 = \rho_{\rm val},
\nonumber \\
\langle \bar{\psi} i \gamma_5  \psi \rangle_+ & = &  \frac{1}{2m} (\partial_x + 2P) \rho_{\rm val},
\label{D6}
\end{eqnarray}
with bound state wave functions labelled by $\ell$. In the second line of Eq.~(\ref{D6}), we have used Eq.~(\ref{D2}).
According to Eq.~(\ref{C7}), the term $\sim \partial_x P$ in (\ref{D4}) can be expressed through the induced density,
\begin{equation}
i \partial_t \psi = \left( - \frac{\partial_x^2}{2m} - \frac{\pi}{N} \left( \rho_{\rm val} - \rho_{\rm ind} \right)\right) \psi.
\label{D7}
\end{equation}
Actually, as a consequence of axial current conservation, the total density $\rho = \rho_{\rm val}+\rho_{\rm ind}$ 
vanishes, so that the two terms in the potential are equal. To see that this also holds in the effective theory, 
take Eq.~(\ref{C3}) to lowest order in the $\epsilon$-expansion,
\begin{equation}
\left( - \partial_x^2 - \frac{4\pi m}{N} \rho_{\rm val}  \right) P = - \frac{2\pi m}{N} (\partial_x + 2P ) \rho_{\rm val}.  
\label{D8}
\end{equation}
The terms without derivatives cancel and we get the constraint
\begin{equation}
\partial_x \left( \partial_x P - \frac{2\pi m}{N} \rho_{\rm val}  \right) = 0.
\label{D9}
\end{equation}
The expression in parentheses is constant. Assuming localized bound states, it must vanish asymptotically, so that the constant is 0, 
\begin{equation}
\partial_x P =   \frac{2\pi m}{N} \rho_{\rm val}.
\label{D10}
\end{equation}
Comparison with Eq.~(\ref{C7}) then shows that $\rho_{\rm val} = - \rho_{\rm ind}$, as expected.
If we plug this into Eq.~(\ref{D7}), we see that the two potential terms are identical and recover the multi-component 
non-linear Schr\"odinger (NLS) equation \cite{17}
\begin{equation}
\partial_t \psi_{1,k} 
 =  \left( - \frac{\partial_x^2}{2m} - 2 \pi \sum_{\ell} \nu_{\ell} |\psi_{1,\ell}|^2 \right) \psi_{1,k} 
\label{D11}
\end{equation}
with $\nu_{\ell} = N_{\ell}/N$.
To reproduce the result of Ref.~\cite{8} for the non-chiral GN model, one can just set $P=0$ in Eq.~(\ref{D4}) and get a similar equation, but with a coupling constant
differing by a factor of 2. Hence the TDHF equations for the massless GN and NJL$_2$ models both reduce to 
the multicomponent NLS equation, to LO in the non-relativistic expansion, with well-known explicit solutions \cite{17}.

\section{Twisted kink in the non-relativistic limit}
\label{sect5}

The result that the NLS equation (\ref{D11}) applies to both the GN and the NJL$_2$ model raises the question: How can one see the expected differences between the
two theories in the non-relativistic limit, for instance the fact that twist and induced fermion density exist only in the NJL$_2$ model? To clarify this 
point, we shall consider the single kink in more detail in this section. 
For a single kink with filling fraction $\nu$, the TDHF equation (\ref{D11}) reduces to
\begin{equation}
i \partial_t \psi_{1}= \left( - \frac{\partial_x^2}{2m}  - 2 \pi \nu |\psi_1|^2  \right) \psi_{1}. 
\label{E1}
\end{equation}
Using 
\begin{equation}
\nu = \frac{\varphi}{\pi},
\label{E2}
\end{equation}
the solution of the NLS equation (\ref{E1}) corresponding to a kink at rest is 
\begin{equation}
\psi_1  =  \sqrt{\frac{m\varphi}{2}} \frac{1}{\cosh \xi} e^{im\varphi^2t/2}
\label{E3}
\end{equation}
with $\xi=m\varphi x$. The energy eigenvalue is
\begin{equation}
E_0 = - \frac{m\varphi^2}{2},
\label{E4}
\end{equation}
the valence fermion density reads
\begin{equation}
\rho_{\rm val} = N_0 \frac{m\varphi}{2}\frac{1}{\cosh^2 \xi}, \quad N_0 = N \nu.
\label{E5}
\end{equation}
These results differ from the exact ones (cf. Sect.~\ref{sect2}) only in that the trigonometric functions $\cos \varphi, \sin \varphi$ have been 
approximated by their Taylor expansions up to O($\varphi^2$). For a boosted kink moving with velocity $v$, there will
be further differences related to kinematics, but this will not be needed here.
The fermion wave functions of the NJL$_2$ and GN models look identical. 
In order to see the twist in the NJL$_2$ model, it is necessary to reconstruct the Dirac potentials $S,P$ first. 
Integration of Eq.~(\ref{D10}) yields
\begin{eqnarray}
P & = & P(-\infty) + \frac{2 \pi m}{N} \int_{-\infty}^x d x'  \rho_{\rm val}(x') .
\label{E6}
\end{eqnarray}
Starting from $P(-\infty) = m \sin \varphi \approx m \varphi$ (to match the choice made in Fig.~\ref{fig1}) and inserting $\rho_{\rm val}$ from (\ref{E5}), we find
\begin{equation}
P = m \varphi \tanh \xi.
\label{E7}
\end{equation}
From this we can determine $S$, using the first line of Eq.~(\ref{C4})
\begin{eqnarray}
S & = & m -  \pi \nu |\psi_1|^2 - \frac{P^2}{2m}
\nonumber \\
& = & m \left( 1 - \frac{\varphi^2}{2} \right) .
\label{E8}
\end{eqnarray}
Combining $S$ and $P$ into the complex potential $\Delta$,
\begin{equation}
\Delta = S-iP = m \left( 1 - \frac{\varphi^2}{2} \ - i \varphi \tanh \xi \right),
\label{E9}
\end{equation}
we recognize the twisted kink for small twist angle, see Eq.~(\ref{B5}) and Fig.~\ref{fig1}, correct to O($\varphi^2$).

Finally, let us compute the mass of the kink in the non-relativistic limit. The mass per fermion 
can be found by taking the energy eigenvalue and adding a double counting correction for the 
potential energy characteristic for the Hartree-Fock (HF) approach,
\begin{equation}
\frac{M}{N_0} =  m+ E_0 +  \int dx  \pi \nu |\psi_1|^4.  
\label{E10}
\end{equation}
Upon inserting $\psi_0$ and $E_0$  from (\ref{E3},\ref{E4}), one finds
\begin{equation}
\frac{M}{N_0}  =  m  \left( 1- \frac{\varphi^2}{6} \right),
\label{E11}
\end{equation}
once again correct to O($\varphi^2$).

\section{Non relativistic limit of the massive NJL$_2$ model}
\label{sect6}
The preceding sections were dealing with the massless NJL$_2$ model. We now turn to the massive model based on the Lagrangian
\begin{equation}
{\cal L}  =  \bar{\psi} \left( i \partial \!\!\!/ - m_0 \right) \psi + \frac{g^2}{2}\left[  (\bar{\psi}\psi)^2 + (\bar{\psi} i \gamma_5 \psi)^2 \right] .
\label{F1}
\end{equation}
The massive NJL$_2$ model contains a new parameter, the bare mass $m_0$. Like the bare coupling constant $g^2$, $m_0$ is not an observable.
In the renormalization process, $g^2$ and $m_0$ get replaced by the physical fermion mass $m$ and a second renormalization group invariant parameter,
commonly chosen as
\begin{equation}
\gamma = \frac{\pi}{Ng^2}\frac{m_0}{m}
\label{F2}
\end{equation}
("confinement parameter"). The vacuum gap equation becomes \cite{4,18} 
\begin{equation}
\frac{\pi}{Ng^2} = \gamma + \ln \frac{\Lambda}{m}.
\label{F3}
\end{equation} 
Due to the explicit breaking of chiral symmetry by the mass term, the vacuum becomes unique ($\Delta = m$).
The pion acquires a mass $\mu$. To leading order in $\gamma$, one finds \cite{19}
\begin{equation}
\mu^2  = - \frac{4 \pi m_0}{N} \langle \bar{\psi} \psi \rangle = 4 m^2 \gamma
\label{F4}
\end{equation}
in full analogy to the Gell-Mann, Oakes, Renner relation in QCD \cite{20}.
As discussed in more detail in Ref.~\cite{9}, the detailed form of the no-sea effective theory depends on the regime.
For the region of small pion masses we are interested in here, the only change in ${\cal L}_{\rm eff}$ of Eq.~(\ref{C1})
is an additional mass term for the pion field,
\begin{equation}
{\cal L}_{\rm eff}(\mu) = {\cal L}_{\rm eff}|_{\rm Eq.(10)} - \frac{1}{2} \mu^2 \Pi^2.
\label{F5}
\end{equation} 
Consequently, the Euler-Lagrange equation (\ref{C2}) for the fermion field is unchanged. The equation for the pion field, Eq.~(\ref{C3}),
gets an extra mass term,
\begin{equation}
\left( \square + \mu^2  - \frac{4\pi m}{N} \bar{\psi}\psi \right) \Pi  =  \sqrt{\frac{4\pi}{N}} m \bar{\psi}i \gamma_5 \psi.
\label{F6}
\end{equation}
As far as the TDHF equation is concerned, we can immediately jump to the non-relativistic reduction where the derivation of 
Eq.~(\ref{D4}) remains valid,
\begin{equation}
i \partial_t \psi_1 = \left( - \frac{\partial_x^2}{2m} - \frac{\pi}{N} \rho_{\rm val} - \frac{(\partial_x P)}{2m} \right) \psi_1.
\label{F7}
\end{equation}
By contrast, the LO equation for $P$, Eq.~(\ref{D8}), is modified by the mass term
\begin{equation}
\left( - \partial_x^2  + \mu^2 - \frac{4\pi m}{N} \rho_{\rm val}  \right) P = - \frac{2\pi m}{N} (\partial_x + 2P ) \rho_{\rm val}  .
\label{F8}
\end{equation}
The terms proportional to $\rho_{\rm val}P$ cancel as before and we get
\begin{equation}
\left( - \partial_x^2 + \mu^2 \right) P = - \frac{2\pi m}{N} \partial_x \rho_{\rm val}.
\label{F9}
\end{equation}
This equation could also have been derived from the PCAC relation for the axial current, no longer
conserved due to the mass term. In order to eliminate $P$ from the TDHF equation (\ref{F7}),
let us solve equation (\ref{F9}) for $P$ formally, 
\begin{equation}
P  =   -  \frac{2 \pi m}{N} \frac{1}{-\partial_x^2+\mu^2} \partial_x \rho_{\rm val}.
\label{F10}
\end{equation}
Then we apply $\partial_x$ to the result,
\begin{equation}
\partial_x P  =   \frac{2\pi m}{N} \left( 1 -  \frac{\mu^2}{-\partial_x^2+\mu^2} \right) \rho_{\rm val}.
\label{F11}
\end{equation}
The first term on the right hand side of Eq.~(\ref{F11}) is familiar from the chiral limit, see Eq.~(\ref{D10}).
The second one is new and has a rather simple physical interpretation. Since
\begin{equation}
\partial_x P = - \frac{2\pi m}{N} \rho_{\rm ind} =  \frac{2\pi m}{N} (\rho_{\rm val}- \rho ),
\label{F12}
\end{equation}
the new term is seen to be proportional to the total fermion density,
\begin{equation}
\rho = \frac{\mu^2}{- \partial_x^2+ \mu^2 } \rho_{\rm val}.
\label{F13}
\end{equation}
Inserting (\ref{F11}) into (\ref{F7}), we arrive at 
\begin{equation}
i \partial_t \psi_1 = \left( - \frac{\partial_x^2}{2m} -  \frac{2\pi}{N} \rho_{\rm val} + \frac{\pi}{N} \frac{\mu^2}{-\partial_x^2+\mu^2} \rho_{\rm val} \right)  \psi_1.
\label{F14}
\end{equation}
Whereas the first interaction term (the same as in the chiral limit) is characteristic for the HF potential of
fermions interacting via a zero range potential, the second one corresponds to a long range Yukawa potential, the 
one-pion exchange potential. 
The non-locality can be eliminated at the cost of getting two coupled nonlinear differential equations
\begin{equation}
i \partial_t \psi_1  =   \left( - \frac{\partial_x^2}{2m} -  \frac{2\pi}{N} \rho_{\rm val} + \frac{\pi}{N} \rho \right)  \psi_1,
\label{F15}
\end{equation}
\begin{equation}
\left( - \partial_x^2 + \mu^2 \right) \rho = \mu^2 \rho_{\rm val}.
\label{F16}
\end{equation}
The non-linearity becomes fully explicit if we recall the form of the valence fermion density,
\begin{equation}
\rho_{\rm val} = \sum_{\ell} N_{\ell} |\psi_{1,\ell}|^2.
\label{F17}
\end{equation}
If we have $K$ bound states, Eq.~(\ref{F16}) has to be written down for all $K$ bound state spinors $\psi_{1,k}$, so that we are dealing with
a system of $K+1$ coupled, nonlinear PDE's.

Once equations (\ref{F15},\ref{F16}) have been solved, the Dirac potentials $S,P$ can be reconstructed as follows
\begin{eqnarray}
P & = & \frac{2\pi m}{N} \int_0^x dx' [\rho_{\rm val}(x')- \rho(x')],
\nonumber \\
S & = & m - \frac{\pi}{N} \rho_{\rm val} - \frac{P^2}{2m}.
\label{F18}
\end{eqnarray}
One would not expect the non-relativistic HF problem with $\delta$-function and Yukawa two-body potentials to be explicitly solvable.
However, provided one works in a regime where the pion Compton wavelength, $\mu^{-1}$, and the scale of the valence fermion density, 
$(m \varphi)^{-1}$, are very different, one can get a lot of insight via analytical approximations. This will be demonstrated
for the single kink at rest in the following section. 

\section{Untwisted kink --- analytical approximations}
\label{sect7}

Consider the (non-relativistic) one-kink problem for the massive NJL$_2$ model. Working in the kink rest frame, we go to the stationary Schr\"odinger equation, choosing a real bound state wave function $\psi$
(we drop the Dirac index of $\psi_1$ from here on). The valence fermion density for occupation fraction $\nu$ reads
\begin{equation}
\rho_{\rm val} = N \nu \psi^2 = N_0 \psi^2.
\label{G1}
\end{equation}
Our task is to solve the coupled set of equations (\ref{F15},\ref{F16}) for $\psi, \rho$, now specialized to
\begin{eqnarray}
\left(  - \frac{\partial_x^2}{2m} - 2  \pi \nu \psi^2 + \frac{\pi}{N} \rho \right) \psi & = & E \psi,
\nonumber \\
\left( - \partial_x^2 + \mu^2 \right) \rho & = & \epsilon \mu^2 N_0 \psi^2 .
\label{G2}
\end{eqnarray} 
In the second line, 
we have introduced a formal expansion parameter $\epsilon$ (different from $\mu$) to 
start a perturbation theory around the chiral limit ($\epsilon=0$). Keeping only terms up to 
first order in $\epsilon$, we set
\begin{equation}
\psi = \psi_0 + \epsilon \psi_1, \quad \rho = \epsilon \rho_1, \quad E = E_0 + \epsilon E_1,
\label{G3}
\end{equation}
where the unperturbed quantities refer to the chiral limit,
\begin{equation}
\left(  - \frac{\partial_x^2}{2m} - 2  \pi \nu \psi_0^2   \right) \psi_0  =  E_0 \psi_0.
\label{G4}
\end{equation}
The results for $\psi_0, E_0$ have been given above, see (\ref{E3},\ref{E4}).
Linearizing the system (\ref{G2}) in $\epsilon$ defines the first correction due to the bare fermion mass as 
\begin{eqnarray}
\left( - \frac{\partial_x^2}{2m} - 6 \pi \nu \psi_0^2 - E_0 \right) \psi_1 & = & \left(E_1  - \frac{\pi}{N}\rho_1\right) \psi_0,
\nonumber \\
\left( - \partial_x^2 + \mu^2 \right) \rho_1 & = & \mu^2 N_0 \psi_0^2.
\label{G5}
\end{eqnarray}
These equations are inhomogeneous, linear differential equations for $\psi_1, \rho_1$. They can be
integrated successively using Green's functions, starting from the 2nd equation. 
This second equation requires only the free Green's function of the Schr\"odinger equation (1d Yukawa potential)
\begin{equation}
\left( - \partial_x^2 + \mu^2 \right) \frac{e^{-\mu |x-x'|}}{2\mu} = \delta(x-x')
\label{G6}
\end{equation} 
and has the result 
\begin{eqnarray}
\rho_1(x) & = &  \frac{\mu}{2} \left[ e^{-\mu x}\int_{-\infty}^x dx' e^{\mu x'} \rho_{\rm val}^0(x') \right.
\nonumber \\
& & + \left. e^{\mu x} \int_x^{\infty} dx' e^{-\mu x'}\rho_{\rm val}^0(x') \right]
\label{G7}
\end{eqnarray}
with $\rho_{\rm val}^0 = N_0 \psi_0^2$.
If we transform the 2nd line of Eq.~(\ref{G5}) to momentum space, we see that
\begin{equation}
\tilde{\rho}_1(k=0) = \tilde{\rho}_{\rm val}(k=0).
\label{G8}
\end{equation}
Hence $\rho_1$ represents a density distribution spread out over distances of O($1/\mu$), as compared to the size of 
$\rho_{\rm val}$ of O($1/\varphi m$), but with the same integrated fermion number. 
One can get an accurate analytic approximation by expanding the exponentials under
the integral into power series and integrating term by term. This generates polylogs in higher order.
Keeping only the LO and NLO terms ($e^{\pm \mu x'} \approx 1 \pm \mu x'$), the result involves only 
elementary functions,
\begin{equation}
\frac{\rho_1}{N_0} = \frac{\mu}{2} \frac{\cosh(\xi-\mu x)}{\cosh \xi} + \frac{\mu^2}{2\varphi m} \left[ \xi \tanh \xi - \ln(2 \cosh \xi) \right],
\label{G9}
\end{equation} 
with $\xi = \varphi m x$. 
\begin{figure}
\begin{center}
\epsfig{file=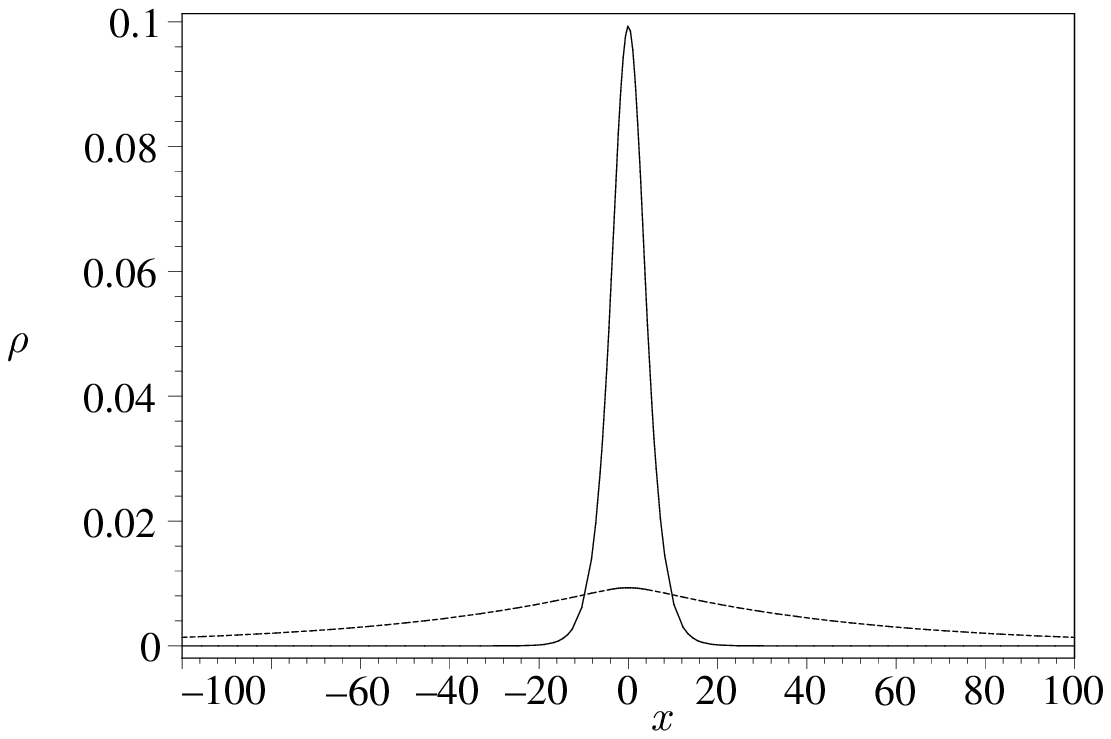,width=8cm,angle=0}
\caption{Valence fermion density (narrow, solid curve) and total fermion density (wide, dashed curve) for the parameters $m=1, \varphi=0.2, \mu=0.02$. 
The broadening is due to the folding with the pion field of size $\sim \mu^{-1}$.}
\label{fig2}
\end{center}
\end{figure}
We illustrate the transition from $\rho_{\rm val}$ to $\rho_1$ by the example in Fig.~\ref{fig2}. Here, we compare the valence fermion density (narrow, solid curve)
with the total density (wide, dashed curve). The drastic difference between these two densities is due to the folding with the pion Green's function of range $\mu^{-1}$. 
The area under both curves is exactly the same. 

Coming back to the result (\ref{G9}), if $\mu \ll m \varphi$,
the first term is the dominant, long range part of O($\mu$), the second one of O($\mu^2$) can be neglected for $\mu \to 0$. 
The leading term can be transformed into
\begin{equation}
\frac{\mu}{2} \frac{\cosh(\xi-\mu x)}{\cosh \xi} = \frac{\mu}{2} \left[ \cosh (\mu x) - \tanh \xi \sinh(\mu x) \right].
\label{G10}
\end{equation}
In order to understand the limit $\mu \to 0$, we re-write it as
\begin{eqnarray}
\frac{\rho_1}{N_0} & \approx & \frac{\mu}{2} \left[ e^{\mu x} \left( \frac{1-\tanh \xi}{2} \right) + e^{-\mu x} \left( \frac{1+\tanh \xi}{2} \right) \right]
\nonumber \\
& \to & \frac{\mu}{2} e^{-\mu |x|} \quad ( \mu \ll \varphi m).
\label{G11}
\end{eqnarray}
In the last step, we have used the fact that the functions $(1 \pm \tanh \xi)/2$ look like step functions if viewed on the scale $\mu^{-1}$.
In this limit, $\rho_1$ is nothing but the static Yukawa field generated by a point charge. Hence the baryon picture which emerges  
is that the fermion density gets distributed over the whole volume of a large pion cloud. The valence fermion density is almost point-like in comparison.
However, no ``quark core" shows up in the fermion density due to the screening of any local charge density in the chiral limit. 
The total fermion number of the bound state is equal to the valence fermion number,
\begin{equation}
N_f = \int dx \rho_1(x) = N_0,
\label{G12}
\end{equation} 
independently of $\mu$. The limit $\mu \to 0$ is evidently subtle, as the fermions get distributed over an infinite volume with vanishing
density. The fact that our state carries fermion number is in conflict with the common lore about the twisted kink as derived starting from $\mu=0$.
Here, the total fermion number has been assumed to vanish, since the fermion density was identically zero. The reason for this discrepancy
is now clear: The extra term in the density behaves like $\mu$ for $\mu \to 0$, but in the total fermion number this is compensated by the
diverging volume ($\sim \mu^{-1}$) over which the charge is spread out. We shall come back to this issue later.

Consider now the first line of equation (\ref{G5}). Its structure becomes more illuminating if we introduce the variable $\xi = m \varphi x$  
and plug in $E_0, \psi_0$ on the left hand side,
\begin{equation}
\left( \partial_{\xi}^2 + \frac{6}{\cosh^2 \xi} -1 \right) \psi_1  = - \frac{2}{m \varphi^2} \left( E_1 - \frac{\pi}{N} \rho_1 \right) \psi_0.
\label{G13}
\end{equation}
This is the (inhomogeneous) $n=2$ P\"oschl-Teller equation, at the energy of the upper bound state.
The regular and irregular solutions (at infinity) of the homogeneous equation can be readily found,
\begin{eqnarray}
f_{\rm reg}(\xi) & = & \frac{\sinh \xi}{\cosh^2 \xi},
\nonumber \\
f_{\rm irreg}(\xi) & = & 3 \xi \frac{\sinh \xi}{\cosh^2 \xi} - \frac{3}{\cosh \xi} +  \cosh \xi ,
\label{G14}
\end{eqnarray}
with Wronski determinant
\begin{equation}
W = f_{\rm reg}'  f_{\rm irreg} -  f_{\rm irreg}'  f_{\rm reg} = -2.
\label{G15}
\end{equation}
The regular solution is the upper bound state wave function, this is the reason why it has a node. 
The Green's function needed to solve Eq.~(\ref{G5}),  
\begin{equation}
\left( \partial_{\xi}^2 + \frac{6}{\cosh^2 \xi} -1 \right) G(\xi,\xi') = \delta(\xi-\xi')
\label{G16}
\end{equation}
is then given by
\begin{equation}
G(\xi, \xi') = - \frac{1}{2} f_{\rm reg}(\xi_>) f_{\rm irreg}(\xi_<)
\label{G17}
\end{equation}
(for $\xi, \xi'>0$).
The right hand side of Eq.~(\ref{G13}) still contains the unknown energy correction $E_1$. This motivates us to first decompose $\psi_1$ according to  
\begin{equation}
\psi_1 = E_1 g_1 + g_2
\label{G18}
\end{equation} 
where 
\begin{equation}
\left( \partial_{\xi}^2 + \frac{6}{\cosh^2 \xi} -1 \right) g_i  = h_i \quad (i=1,2)
\label{G19}
\end{equation}
and
\begin{eqnarray}
h_1 & = & - \frac{2}{m\varphi^2} \psi_0,
\nonumber \\
h_2 & = & \frac{2}{m \varphi^2} \frac{\pi}{N} \rho_1 \psi_0.
\label{G20}
\end{eqnarray}
The solutions of the inhomogeneous equations (\ref{G19}) are now independent of $E_1$ and given by 
\begin{eqnarray}
g_i(\xi ) & = &  - \frac{1}{2}  \left[ f_{\rm reg}(\xi ) \int_0^{\xi} d\xi' f_{\rm irreg}(\xi') h_i (\xi') \right. 
\nonumber \\
&  &    \left. +  f_{\rm irreg}(\xi) \int_{\xi}^{\infty} d\xi' f_{\rm reg}(\xi') h_i (\xi') \right]. 
\label{G21}
\end{eqnarray}
$g_1$ can be computed analytically. In the case of $g_2$, we get an accurate analytical approximation by expanding the function $h_2$ in the integrand in powers
of $\mu$ and truncating the expansion at O($\mu^2$), consistent with our 
effective Lagrangian. The energy $E_1$ can then be found afterwards from the orthogonality condition between $\psi_0$ and
$\psi_1$, a necessary condition for the normalization of $\psi$ to O($\epsilon$),
\begin{equation}
\int_{-\infty}^{\infty} dx \psi_0(x) \psi_1(x) = 0.
\label{G22}
\end{equation} 
The result for $E_1$ is
\begin{equation}
E_1 = \frac{\varphi \mu}{2} - \frac{\mu^2}{4m},
\label{G23}
\end{equation}
whereas we find the following expression for $\psi_1$ to O($\mu^2$), 
\begin{eqnarray}
\psi_1 & = & - \frac{\mu^2 \sqrt{2m\varphi}}{4 m^2 \varphi^2} \tilde{\psi}_1,
\nonumber \\
\tilde{\psi}_1 & = & \frac{\ln ( 2 \cosh \xi )}{\cosh \xi}  
 + \frac{1}{4} \frac{\sinh \xi}{\cosh^2 \xi} \left[ {\rm dilog}\left(e^{2\xi}+1\right) \right.
\nonumber \\
& & - \left. {\rm dilog}\left( e^{-2\xi}+1 \right) - 4 \xi \right].
\label{G24}
\end{eqnarray} 
Next we compute the Dirac potentials $S,P$. We first have to 
integrate the equation for $\partial_xP$,
\begin{equation}
P = \frac{2 \pi m}{N} \int_0^x dx' \left[ \rho_{\rm val}(x') - \rho_1(x') \right]. 
\label{G25}
\end{equation}
For consistency, $\rho_{\rm val}$ has to be treated to first order in $\epsilon$,
\begin{equation}
\rho_{\rm val} = N_0 \left( \psi_0^2 + 2 \psi_0 \psi_1 \right).
\label{G26}
\end{equation}
Here it would be wrong to simply expand the integrand in powers of $\mu$. In order to treat the ``soft" part of $\rho_1$ correctly,
we have to do one partial integration with respect to the slowly varying exponentials $e^{\pm \mu x}$ first.
The remainder of the integrand can then be expanded into a power series in $\mu$ and truncated at O($\mu^2$).
The result for $P$ is
\begin{eqnarray}
P & = & P_{\rm soft} + P'
\label{G27} \\
P_{\rm soft} & = & m \varphi \frac{ \sinh (\xi - \mu x)}{\cosh \xi},
\nonumber \\
P' & = & \frac{\mu^2}{8m\varphi} \left[ \left( \frac{1}{\cosh^2 \xi} - 2 \right) \left( {\rm dilog}(e^{2\xi}+1) \right. \right.
\nonumber \\
& & \left. - {\rm dilog}(e^{-2\xi}+1) \right)
\nonumber \\
& & - 4 \left. \tanh \xi \left( \xi^2 + \ln (2 \cosh \xi)\right) + 4 \xi \tanh^2 \xi \right].
\nonumber 
\end{eqnarray}
As in the case of the fermion density $\rho_1$, we can infer the asymptotics of $P$ for $\mu \to 0$ by
expanding the trigonometric function in the numerator of $P_{\rm soft}$,
\begin{equation}
m\varphi  \frac{\sinh (\xi - \mu x)}{\cosh \xi} = m \varphi \left[ \cosh(\mu x) \tanh \xi - \sinh(\mu x) \right].
\label{G28}
\end{equation}
Asymptotically, for $\mu \to 0$, $P'$ can be neglected so that 
\begin{eqnarray}
P & \approx & m\varphi \left[ - e^{\mu x} \left( \frac{1-\tanh \xi}{2} \right) + e^{-\mu x} \left( \frac{1+\tanh \xi}{2} \right) \right]
\nonumber \\
& \to & m \varphi \epsilon(x) e^{-\mu |x|}
\label{G29}
\end{eqnarray}
with $\epsilon(x)$ the sign function.
\begin{figure}
\begin{center}
\epsfig{file=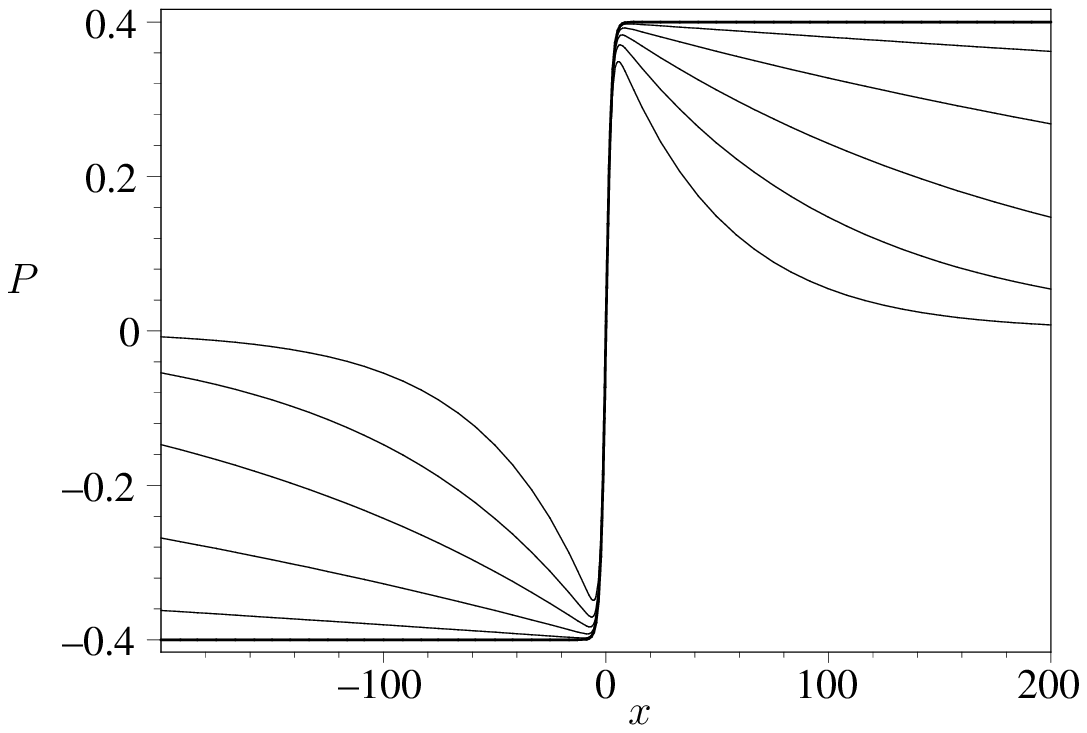,width=8cm,angle=0}
\caption{Pseudoscalar potential $P$ as a function of $x$. The different curves correspond to various values of $\mu$ and show the approach to the
$\tanh$-shape in the chiral limit (thick line). Parameters: $m=1, \varphi=0.4, \mu=0.02,0.01,0.005,0.002,0.0005$. The larger $\mu$, the faster $P$ goes to 0 asymptotically.} 
\label{fig3}
\end{center}
\end{figure}
Fig.~\ref{fig3} shows how $P$ approaches the chiral limit with decreasing pion mass. For any finite $\mu$, $P$ vanishes for $x \to \pm \infty$, whereas it goes to  
$\pm \varphi$ for $x \to \pm \infty$ at $\mu=0$. Given $P$ and $\psi_1$, the scalar mean field $S$ can be computed from
\begin{equation}
S = m - \frac{\pi}{N} \rho_{\rm val}  - \frac{P^2}{2m}.
\label{G30}
\end{equation}
\begin{figure}
\begin{center}
\epsfig{file=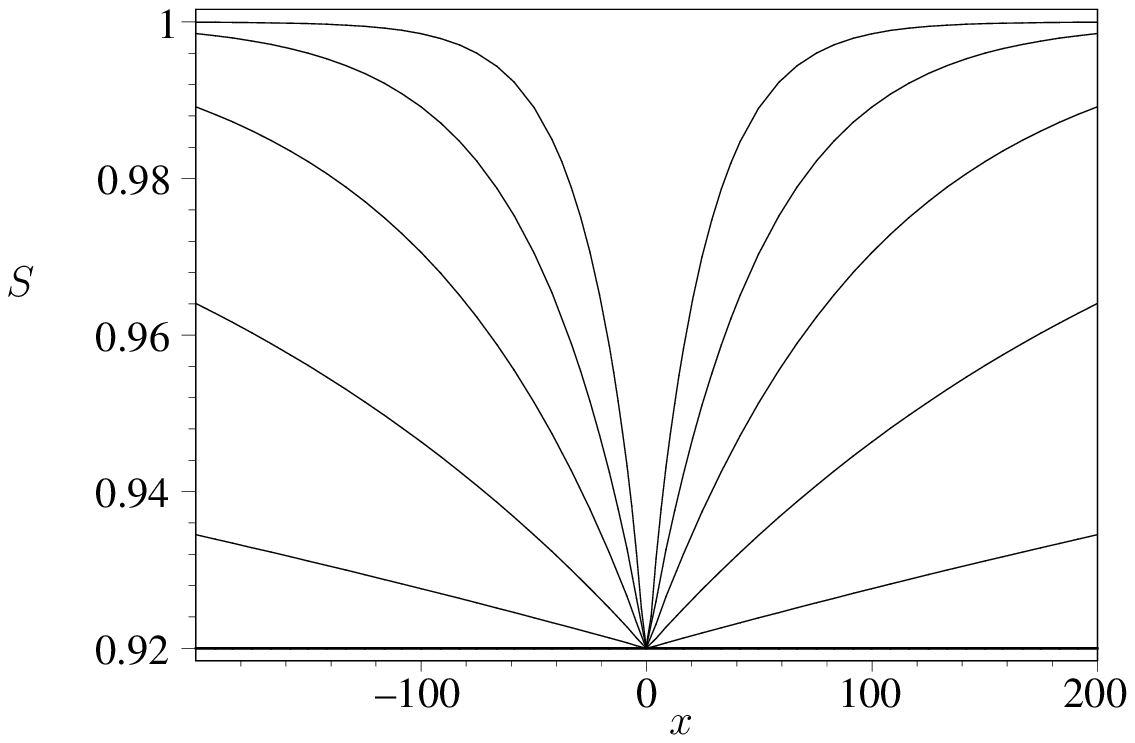,width=8cm,angle=0}
\caption{Same as Fig.~\ref{fig3}, but for scalar potential $S$. The larger $\mu$, the faster $S$ goes to 1 asymptotically. In the chiral limit,
$S$ has the constant value $m(1-\varphi^2/2)$ (thick line).}
\label{fig4}
\end{center}
\end{figure}
Fig.~\ref{fig4} shows the evolution of the scalar potential with $\mu$, for the same parameters as in Fig.~\ref{fig3}.
$S/m$ has the constant value ($1-\varphi^2/2$) in the chiral limit,  but has to go to 1 asymptotically for any non-zero $\mu$.

\begin{figure}
\begin{center}
\epsfig{file=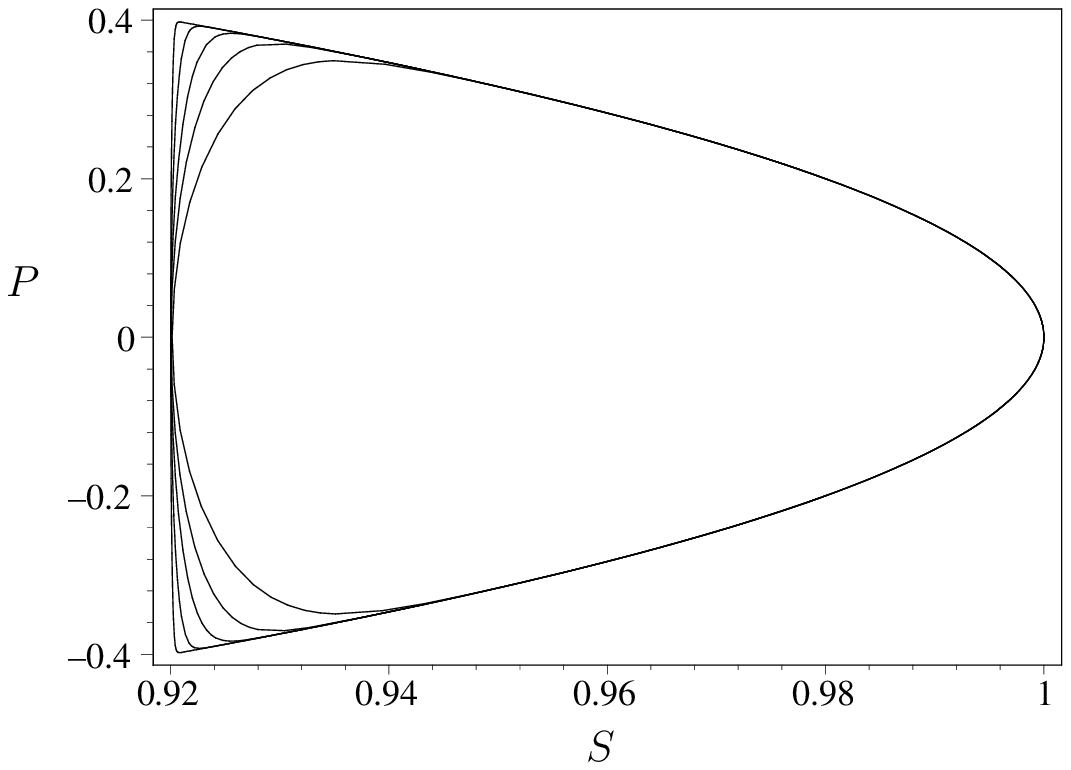,width=8cm,angle=0}
\caption{Contour plot in the ($S,P$)-plane for the same parameters as used in Figs.~\ref{fig3},\ref{fig4}. Different scales had to be used on the $S$- and 
$P$-axes to exhibit the evolution of the closed curves with decreasing $\mu$. The innermost curve belongs to the largest pion mass.}
\label{fig5}
\end{center}
\end{figure}
\begin{figure}
\begin{center}
\epsfig{file=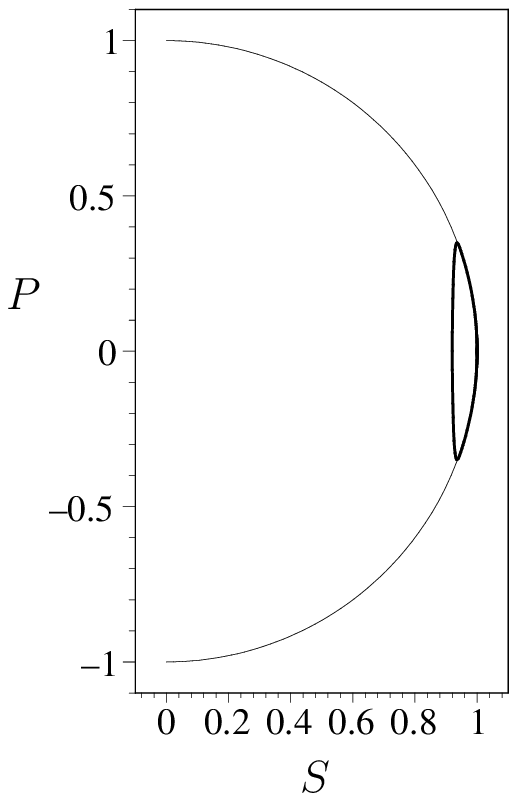,width=4cm,angle=0}\epsfig{file=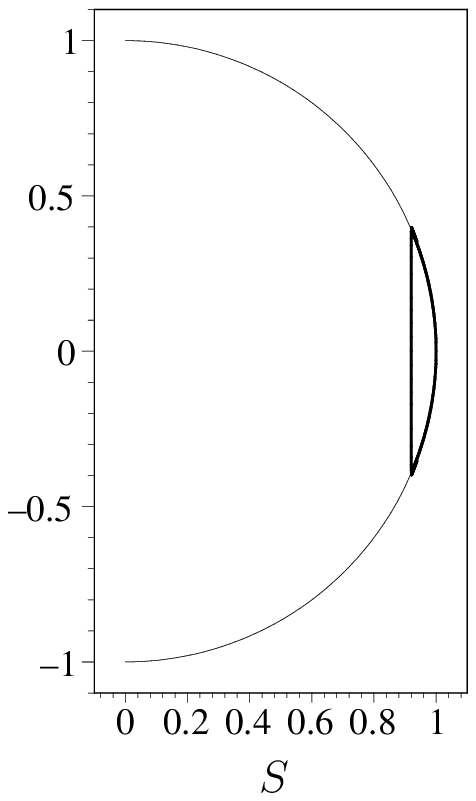,width=3.7cm,angle=0}
\caption{Similar kind of plots as in Fig.~\ref{fig5}, but using the same scale on both axes. Left hand curve: $\mu=0.02$, right hand curve: $\mu=0.0005$.
The chord of the twisted kink emerges neatly, but the contour is closed along the chiral circle for any non-zero pion mass.}
\label{fig6}
\end{center}
\end{figure}

The $(S,P)$ plot is perhaps more instructive. Fig.~\ref{fig5} shows such contour plots for the parameters of Figs.~\ref{fig3}, \ref{fig4}.
We have used different scales on the $S$ and $P$ axes for better visibility of the details. What actually happens becomes clearer if we
plot $S$ and $P$ on the same scale, see Fig.~\ref{fig6}. The left (right) curve corresponds to the innermost (outermost) curve of Fig.~\ref{fig5}.  
As $\mu$ is sent to 0, the chord characteristic of the twisted kink becomes very conspicuous. However, the curve in the ($S,P$) plane must now be closed at the unique vacuum 
point ($S=m,P=0$). This is apparently achieved by closing the contour along the arc of the chiral circle joining the endpoints of the chord. This confirms earlier 
indications of a discrepancy between the results reached by setting $\mu=0$ from the beginning and by approaching the point $\mu=0$ from above.
The question whether this corresponds to a difference in the physics will be discussed in the next section.

For completeness, we have also computed the baryon mass as a function of $\mu$. The HF double counting correction is now given by 
\begin{equation}
\frac{\Delta E_{\rm d.c.}}{N} = \nu  \int dx \left(    \varphi \psi^4 - \frac{\varphi}{2} \psi^2 \frac{\mu^2}{-\partial_x^2+\mu^2} \psi^2  \right),
\label{G31}
\end{equation}
to be evaluated to first order in $\epsilon$ only.
The baryon mass can then be obtained from
\begin{equation}
\frac{M}{N} = \nu ( m + E) +  \frac{\Delta E_{\rm d.c.}}{N}.  
\label{G32}
\end{equation}
Evaluating the double counting correction to O($\mu^2$), we find
\begin{equation}
\frac{\Delta E_{\rm d.c.}}{N}  = \nu \left( \frac{\varphi^2 m}{3} + \frac{\varphi \mu}{4} - \frac{\mu^2}{2m}\right)
\label{G33}
\end{equation}
and thus the baryon mass becomes
\begin{equation}
\frac{M_B}{N}   = \nu m \left[ 1 - \frac{\varphi^2}{6} + \frac{3 \varphi}{4} \frac{\mu}{m} - \frac{3}{4} \left( \frac{\mu}{m} \right)^2 \right].
\label{G34}
\end{equation}

\section{Comparison to NJL$_2$ baryons with maximal fermion number}
\label{sect8}

To put our results into perspective, it is worthwhile to recall what else is known about NJL$_2$ baryons near the chiral limit.
In the present work, we have studied the non-relativistic region of small filling fraction ($N_0 \ll N$) and small or vanishing pion masses.
Analytical results could be obtained owing to the assumption that three length scales are well separated,
\begin{equation}
\frac{1}{\mu} \gg \frac{1}{m \varphi} \gg \frac{1}{m}.
\label{H1}
\end{equation} 
\begin{figure}
\begin{center}
\epsfig{file=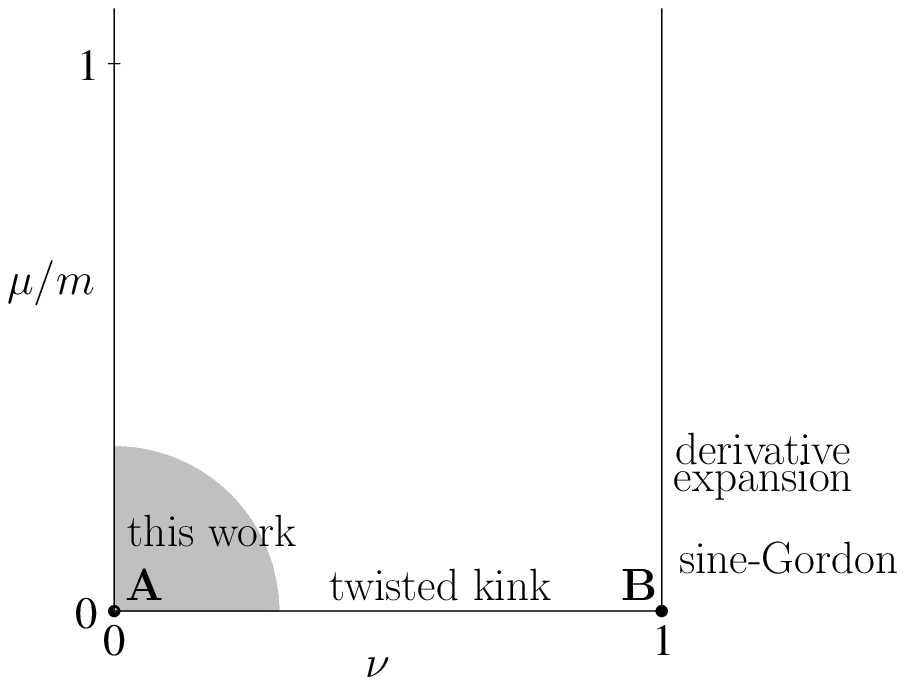,width=8cm,angle=0}
\caption{Map of what is known about baryons in the massive NJL$_2$ model, in the vicinity of the chiral limit. See main text for 
a discussion.}
\label{fig7}
\end{center}
\end{figure}
The first inequality was instrumental when computing certain integrals in closed form, the second one is a precondition for being able to apply 
the no-sea effective theory and the non-relativistic reduction. In Fig.~\ref{fig7}, we show what else is known in the ($\nu, \mu$) plane. The twisted kink of Shei lives on the $\nu$-axis ($\mu=0$),
in the whole interval $[0,1]$. The $\mu$-axis ($\nu=0$) corresponds to vacua. At the point $\mu=0$ (point A), the vacua are degenerate and lie on the chiral
circle ($\Delta = m e^{i\theta}$). For any $\mu>0$, the vacuum is unique ($\Delta=m$). The shaded region around point A is the topic of the present work.
The other region which has been explored before is the vertical line $\nu=1$ of fermion number $N$. Here, for small $\mu>0$, the kink can be mapped quantitatively 
onto the kink of the sine-Gordon equation, as shown by Salcedo et al. \cite{10}. The resulting mean field is 
\begin{equation}
\Delta  =  e^{2i \chi}, \quad
\chi  =   2 \arctan e^{\mu x},
\label{H2}
\end{equation}
the baryon mass 
\begin{equation}
M_{\rm B} = N \frac{2\mu}{\pi}. 
\label{H3}
\end{equation}
The fermion density is proportional to $\partial_x \chi$, 
\begin{equation}
\frac{\rho}{N} = \frac{1}{\pi} \partial_x \chi = \frac{\mu}{\pi \cosh (\mu x)},
\label{H4}
\end{equation}
and winding number equals baryon number $N_{\rm B}=N_f/N$.
\begin{figure}
\begin{center}
\epsfig{file=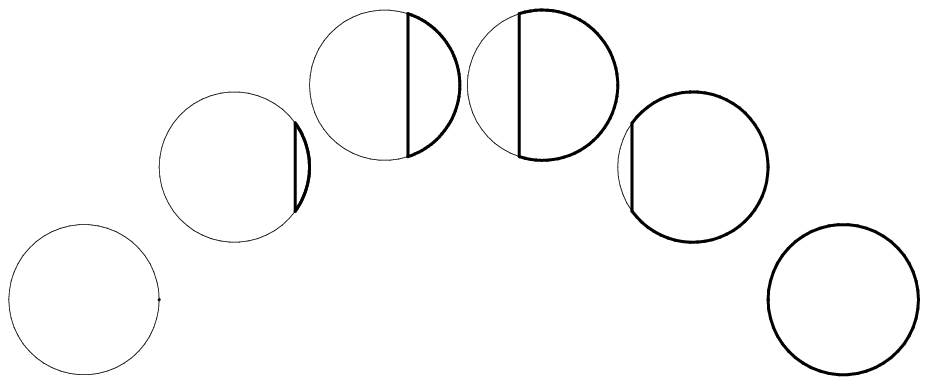,width=8cm,angle=0}
\caption{Conjectured interpolation between the kink baryon with small fermion number (this work) and maximal fermion number
(one-turn chiral spiral). The horizontal and vertical displacement of the circles is a measure for fermion number and mass, respectively.}
\label{fig8}
\end{center}
\end{figure}
For larger $\mu$, systematic corrections to the modulus and phase of $\Delta$ are available from the derivative expansion, apparently converging up to values of $\mu$ 
as large as the fermion mass $m$ \cite{5}. As one descends the line $\nu=1$ towards the point B, the size of the baryon increases like the pion Compton
wavelength $\mu^{-1}$ whereas its mass decreases like $\mu$. At the endpoint B ($\nu=1, \mu=0$) we are left with a completely delocalized, massless 
baryon. In Ref.~\cite{21}, this object has been interpreted in a finite volume of length $L$ as one turn of a ``chiral spiral" with characteristics
\begin{equation}
\Delta = e^{2i\pi x/L}, \quad \frac{\rho}{N}= \frac{1}{L}.
\label{H5}
\end{equation}
What we learn from this is that there are two ways of making sense out of the object located at point B, either by approaching it from finite $\mu$
or from finite volume. Had we set $L=\infty$ and $\mu=0$ from the outset, there would be no way of ``seeing" the fermion number $N$, since the density vanishes
identically.

The reason why we recall this fact should now be obvious. The situation at $N_f=N$ is strongly reminiscent of what we have found at $N_f \ll N$ in the present
work. When comparing the non-relativistic twisted kink as known from the chiral limit with the baryon at $\mu \to 0^+$, we were also facing a 
vanishing fermion density, but a constant fermion number. That these two facts are related can also be seen as follows: If we approach the point B
along the $\nu$ axis, we have to conclude that the states at A and B are vacua (spatially constant, vanishing fermion number.) If we approach 
B along the line $\nu=1$, we conclude that there is a delocalized, massless baryon with fermion number $N$, in contrast to the vacuum at point  A.
In our opinion, the second alternative is the correct one. This is a strong hint that twisted kinks carry fermion number.

Although we have only firm results near the endpoints A and B of the twisted kink interval, we propose the following tentative solution to the 
observed discrepancy between the fermion numbers of twisted and untwisted kinks. Starting from the vacuum at A and moving along the $\nu$ axis, we should 
close the chord of the twisted kink by an arc along the chiral circle, as we have seen at $\mu \to 0^+$. In the chiral limit we now interpret this arc simply 
as a fractional turn of the chiral spiral. In this case, the twisted kink keeps the mass commonly attributed to it, but it carries fermion number arising from the
arc with value
\begin{equation}
N_f(\varphi) = \frac{\varphi}{\pi} N.
\label{H6}
\end{equation} 
In the chiral spiral, fermion number is trivially proportional to the length of the arc. The self-consistency condition of the twisted kink
identifies this fermion number with the valence fermion number, so that we are back at the situation of the GN model where
total and valence fermion number agree with each other. Energy density is located along the chord, fermion number along the arc, but these
two observables are spatially separated in a somewhat counter-intuitive manner. Neither the chord nor the arc can stand alone because there is nothing to stabilize
their end points, but the composite of a twisted kink and a partial turn of the chiral spiral would seem to be a viable, stable state.  
 
This picture is summarized in Fig.~\ref{fig8}. The horizontal position of the circles is proportional to fermion number and ranges from 0 (at the left end) to $N$
(at the right end). The vertical position is proportional to the mass of the corresponding baryons ($\sim \sin \varphi$).
Mass depends only on the length of the chord, so that baryons with $N_f$ and $N-N_f$ fermions are degenerate. Fermion number depends only on the length of the arc,
increasing linearly from left to right. Note that the same picture holds for baryons and anti-baryons, depending on whether the closed
contours are traced out counter-clockwise (baryons) or clockwise (anti-baryons) with increasing $x$.

This whole scenario is still somewhat speculative, since we have quantitative results only at $\nu \ll 1$ and at $\nu=1$ so far. To check whether it is also true
inbetween these extremes would require either an extension of the derivative expansion to partially filled orbits, or a careful finite volume analysis. We leave this
interesting problem to the future.

\section{Summary and conclusions}
\label{sect9}

Twisted kinks of the massless NJL$_2$ model have received a lot of attention in recent years. The motivation of the present study was the desire to better 
understand these unfamiliar objects. To this end, we tried to approach the chiral limit from the side of the massive NJL$_2$ model, where twisted kinks
cannot exist. Since integrability of the NJL$_2$ model is lost for finite bare quark masses, it is a non-trivial task to get analytical insight. We therefore 
concentrated on the non-relativistic regime of weak occupation of the valence level. There, a no-sea effective theory enabled us to reduce the
problem to a simpler, non-relativistic HF problem. As a warm-up, we confirmed the known results in the chiral limit, starting from the effective 
theory for the massless NJL$_2$ model. As soon as one switches on the bare mass term, a long range pion exchange
potential appears in the HF problem, in addition to the usual zero-range potential. This results in a re-destribution of the valence fermion density over the whole volume of the 
pion cloud, the total fermion number remaining the same. In the chiral limit, the fermion density vanishes but the volume diverges, with non-zero
integrated density. This is the reason why one misses fermion number when deriving the twisted kink  from the outset at zero bare fermion mass. 

Taken together with what was already known about baryons with maximal fermion number near the chiral limit, a novel interpretation of twisted 
kinks emerges. We would like to view them as composites of a chord soliton (the standard twisted kink) and a fractional winding of the one-turn
chiral spiral, invented to explain massless baryons with fermion number $N$. Surprisingly, these objects exhibit a spatial separation
of energy density (in the chord) and fermion density (in the arc).

These results remind us that the massless NJL$_2$ model suffers from infrared (IR) problems, an issue which already plagued the 
derivation of the no-sea effective theory. The reason is the massless pion field. A careful analysis should therefore start from
an IR regulated setting, either by enclosing the system into a finite box, or by working at finite bare masses. It will be interesting to see
whether this is feasible for arbitrary fermion numbers, and what are the consequences for problems involving several interacting, twisted kinks.
\newpage

\end{document}